# Evidence for rare-region physics in the structural and electronic degrees of freedom of the nickelate $La_{2-x}Sr_xNiO_4$

R. J. Spieker[1], B. Krohnke[1], D. Zhai[1], A. Lopez Benet[1], M. Spaić[2], X. He[1], C. Y. Tan[1], Z. W. Anderson[3], F. Ye[4], H. Cao[4], M. J. Krogstad[5], R. Osborn[3], D. Pelc[1,2,*], and M. Greven[1,*]

[1]School of Physics and Astronomy, University of Minnesota, Minneapolis, MN 55455, USA

[2]Department of Physics, Faculty of Science, University of Zagreb, 10000 Zagreb, Croatia

[3]Materials Science Division, Argonne National Laboratory, Lemont, 60439 IL, USA

[4]Neutron Scattering Division, Oak Ridge National Laboratory, Oak Ridge, 37831 TN, USA

[5]Advanced Photon Source, Argonne National Laboratory, Lemont, 60439 IL, USA

*Correspondence to: dpelc@phy.hr, greven@umn.edu

**We present a diffuse neutron and x-ray scattering study of structural as well as spin- and charge-density-wave fluctuations in the electrical insulator $La_{2-x}Sr_xNiO_4$. This lamellar nickelate is an isostructural analogue of the high-temperature cuprate superconductor $La_{2-x}Sr_xCuO_4$, for which recent experiments uncovered evidence for unusual structural and superconducting fluctuations indicative of rare-region physics due to inherent inhomogeneity unrelated to common point disorder effects. Here we find closely analogous nanoscale orthorhombic fluctuation behavior in $La_{2-x}Sr_xNiO_4$, including exponential scaling of the diffuse scattering intensity and power-law scaling of the characteristic length with relative temperature. Moreover, our neutron and x-ray scattering data reveal similar behavior for short-range magnetic and charge fluctuations above the respective ordering temperatures. These observations indicate that rare-region effects are a generic feature of perovskite-related structures and lead to universal fluctuations of both structural and electronic degrees of freedom over extended temperature ranges.**

## I. Introduction

Soon after the discovery of superconductivity in the cuprates [1], the nickelates were explored as some of their most promising electronic analogues [2,3]: both materials families have perovskite-derived structures with transition-metal-oxygen layers and exhibit strong electronic correlation effects, including antiferromagnetism and stripe order [4-15]. Despite these similarities, the nickelates are generally insulators, not metals or high-$T_c$ superconductors. However, the recent finding of superconductivity in thin films of the "infinite-layer" compound $Nd_{1-x}Sr_xNiO_2$ [16] triggered a resurgence of interest in the nickelates [17]. Since then, superconductivity has been reported in thin films of quintuple-layer $Nd_6Ni_5O_{12}$ [18], freestanding heterostructures of $Nd_{1-x}Sr_xNiO_2$ [19], and bulk single crystals under hydrostatic pressure [20-23]. Given the extensive similarities between nickelates and cuprates, it is therefore of considerable importance to perform comparative investigations and to obtain deeper insight into the physical mechanisms that underpin their unconventional behavior.

In the case of the cuprates, increasing evidence points to a prevalence of inherent structural and electronic inhomogeneity [24-39]. One of the outstanding features is robust exponential scaling of superconducting fluctuations with relative temperature, $T$ - $T_c$, in stark contrast with power-law scaling with reduced temperature, $(T - T_c)/T_c$, expected for a second-order phase transition [32-34]. Instead, the observed behavior is consistent with the presence of rare regions that are locally ordered, and that result from spatially inhomogeneous fluctuations of the underlying interactions [40-43]. More recently, analogous exponential scaling was observed for short-range *structural* fluctuations in $La_{2-x}Sr_xCuO_4$ (LSCO) and $Tl_2Ba_2CuO_{6+\delta}$ [37,39]. Rigid rotation of $CuO_6$ octahedra causes these materials to undergo a structural transition from a high-temperature tetragonal (HTT) to a low-temperature orthorhombic (LTO) phase at a doping-dependent temperature, $T_{LTO}$ [44] (Fig. 1). Nanoscale LTO fluctuations persist within the tetragonal phase in wide doping and temperature ranges [45-47] and exhibit robust exponential scaling [37,39]. This is indicative of the nucleation of rare nanoscale ordered regions at temperatures significantly above the respective transition temperatures, triggered by a coupling of both structural and superconducting order parameters to some universal, underlying inhomogeneity.

In this work, we combine diffuse x-ray and neutron scattering to study structural and electronic fluctuations in $La_{2-x}Sr_xNiO_4$ (LSNO), with the goal to determine if rare-region effects are present in the nickelates as well. LSNO is an ideal candidate for such an investigation, as it is isostructural with LSCO and shows the same structural transition [48]. On the other hand, the two lamellar oxides exhibit different electronic properties: while bulk charge-spin stripe order is absent in LSCO, the insulator LSNO shows robust stripe order across much of its temperature-doping phase diagram (Fig. 1a). We find that the structural fluctuation behavior of LSNO is nearly indistinguishable from that of LSCO [37,39] and that the spin and charge fluctuations exhibit exponential temperature dependences as well. These results demonstrate that both families of oxides share the same non-trivial underlying inhomogeneity, and they extend the observation of the unusual scaling behavior to two additional electronic order parameters. It is thus a distinct possibility that such behavior is generic and independent of system details, analogous to critical fluctuations close to a second-order phase transition, but prevalent over wide temperature ranges and far from the critical point.

Diffuse scattering provides a wealth of information about structural and electronic correlations, and recent technical developments have resulted in important insights for a wide range of quantum materials [49]. We use x-ray scattering to probe the structural and charge fluctuations in LSNO single crystals with $x = 0.16$, 0.20, and 0.33 at beamline 6-ID-D of the Advanced Photon Source, Argonne National Laboratory. Such measurements naturally involve energy-integration of the dynamic structure factor and hence do not provide information regarding the relevant energy scales. We employ neutron scattering to probe structural and magnetic fluctuations in single crystals with $x = 0.16$ and $x = 0.24$ with the CORELLI spectrometer [50] at the Spallation Neutron Source, Oak Ridge National Laboratory, which is capable of simultaneously capturing the total (energy-integrated to ~10 meV) and quasistatic (< 2 meV) responses.

This paper is structured as follows: Sec. II details the experimental methods used: sample preparation, neutron and x-ray scattering, and data analysis; Sec. III describes our results; in Secs. IV and V, we then discuss and summarize our findings. Additional information is presented in four appendices: Appendix A: Intensity and characteristic length extraction; Appendix B: Spin fluctuations; Appendix C: Temperature scales; Appendix D: Temperature dependences.

## II. Experimental Methods

### A. Sample Preparation

Following the preparation of polycrystalline starting material via solid-state reaction, single crystals of $La_{2-x}Sr_xNiO_4$ ($x = 0.16$, 0.20, 0.24, and 0.33) with masses up to 2 g were grown with the travelling-solvent floating-zone method. Laue x-ray diffraction was used to verify the high structural quality of the crystals. As-grown LSNO tends to contain excess oxygen, and while for $x = 0$ this is known to result in additional structural and electronic phases, the material becomes progressively less sensitive to excess oxygen with increasing Sr concentration [48,51]. The crystals were annealed at 1050°C under vacuum (~3 mTorr) to relax internal stress due to the growth process and to minimize the excess oxygen content.

### B. Neutron and x-ray scattering

Diffuse x-ray scattering measurements using high-energy 87 keV photons were carried out at beamline 6-ID-D of the Advanced Photon Source, Argonne National Laboratory. A fast CdTe detector was used to collect data during full $\phi$ rotations from 0° to 360°, with $\omega$ = -15°, 0°, and 15°, which enabled access to a large reciprocal space volume. X-ray data were analyzed using NXRefine [52].

Diffuse neutron scattering measurements were carried out at the CORELLI spectrometer [50] at the Spallation Neutron Source, Oak Ridge National Laboratory. A white neutron beam (0.7 Å < $\lambda$ < 2.86 Å) was used in conjunction with correlation choppers, which enabled the measurement of relatively small (~250 mg) samples and the simultaneous determination of the total (energy-integrated to ~10 meV) and quasistatic (< 2 meV) responses.

### C. Data analysis

In reciprocal space, features associated with LTO distortions appear at half-integer Bragg positions of the tetragonal structure. These LTO superstructure intensities were extracted following the procedure in ref. [37]; see Fig. 5 in Appendix A for typical data before and after background subtraction. The intensity was obtained by integrating over a cuboid with linear dimensions ~ 0.6 r.l.u. centered on the diffuse LTO features. The size of the integration box was chosen to be sufficiently larger than the largest reciprocal-space width of the relevant diffuse features, and small changes of the box size did not affect the results. Except for the $x$ = 0.24 sample, background correction was performed by subtracting the integrated intensity of an equivalent cuboid centered on a nearby, featureless region of reciprocal space. In the case of the $x$ = 0.24 sample, the background contained significant powder-ring contributions (see Fig. 2(a)), and nested cuboids were used to extract the intensity, with the innermost cuboid corresponding to the LTO feature, and the intensity difference between cuboids giving an estimate of the background. The neutron scattering data in Figs. 2 and 3 correspond to the (2.5 2.5 4) and (3.5 3.5 2) LTO peaks for $x$ = 0.16 and 0.24, respectively. The quasistatic response was multiplied by a factor of ~1.4 for comparison with the total scattering response (see Fig. 2(d)). This normalization factor was extracted by comparing the relative intensities of fundamental Bragg peaks in the quasistatic and total scattering channels, which should converge as the temperature approaches zero. The x-ray data in Figs. 3 and 4 correspond to the $(4.5\ \overline{6.5}\ 2)$, $(\overline{2.5}\ \overline{8.5}\ 2)$, and $(4.5\ \overline{6.5}\ 2)$ LTO peaks for $x$ = 0.16, 0.2, and 0.33, respectively.

The widths of the diffuse LTO features were extracted as follows: line cuts along [100] were fit to a one-dimensional gaussian profile plus a third-order polynomial background at each temperature, after integrating out the intensities along [010] and [001] (integration range ± 0.3 r.l.u. centered at the LTO peak). We systematically varied the integration window and fit range and found that this did not significantly affect the results. One-dimensional line cuts were used, since the diffuse scattering is isotropic in the $HK$-plane, consistent with previous observations for LSCO at higher doping levels [37] (see Fig. 6 in Appendix A for representative data and corresponding fits). Effective characteristic lengths, $\xi$, were estimated using the relation $\xi = 0.41/\sigma$, where the numerical factor was determined from a diffraction calculation for a planar LTO patch with linear size $\xi$ [37].

A somewhat different approach was used to determine spin-order (SO) and charge-order (CO) intensities from the neutron and x-ray scattering data, respectively. The former were obtained by integrating the intensities over rectangular $HK$-plane boxes (linear dimensions of ~ 0.1 r.l.u.) around the artifact-free incommensurate $(0.5\pm\delta,\ \pm0.5\pm\delta, L)$ positions, and by ~8 r.l.u. along [001] over the rods of two-dimensional scattering (the two-dimensional nature of the SO scattering is shown in Fig. 7 in Appendix B). The CO intensities were extracted from the x-ray data by integrating over rectangular boxes around many incommensurate (*HK*0) peaks and then averaging the results. We note that CO was also visible in our neutron scattering data as result of coupling between charge modulations and atomic displacements. However, relatively poor signal-to-noise ratio and limited coverage in the *HK*0 plane prevented an accurate determination of the temperature dependence. Background subtraction for both the CO and SO signals was performed by subtracting the intensities from equivalent boxes centered on nearby, featureless regions of reciprocal space. Streaking due to Compton scattering in the CdTe sensor layer prevented reliable normalization of the x-ray CO signal by nearby Bragg peaks.

## III. Results

Figure 2(a) shows representative scattering data (the neutron total scattering intensity in the *HK*2 scattering plane for $x = 0.24$ at 25 K), whereas Fig. 2(b) summarizes the temperature dependence of the strength of the LTO fluctuations (x-ray and neutron total scattering data). At low temperatures, the $x = 0.16$ and 0.2 samples are in the long-range ordered LTO phase. Above the respective transition temperatures, the intensity of the diffuse features at the LTO wavevectors decreases exponentially as $\exp(-T/T_0)$ with increasing temperature. Although the $x = 0.24$ and 0.33 compositions have average tetragonal structure at all temperatures, LTO fluctuations are robustly present even well above room temperature. The decay constant $T_0$ lies in the 200-300 K range for all samples, comparable to values for LSCO (see Fig. 8 and Table 1 in Appendix C). Figure 2(c) shows the temperature dependences of the characteristic lengths, which were extracted in the same manner as in [37]. Figures 2(d,e) compare the quasistatic and energy-integrated neutron scattering intensities for $x = 0.16$ and 0.24. The LTO fluctuations are seen to decay exponentially in both scattering channels, although $T_0$ of the quasistatic component is about a factor of two smaller (see Table 1 in Appendix C). Therefore, even though the response is increasingly dynamic, there remain considerable static structural distortions at high temperatures.

These findings mirror recent results for the cuprate LSCO [37,39], but there exist subtle differences. In both oxides, the strength of the LTO fluctuations decays exponentially (Fig. 3(a)). Moreover, the LSNO data show the same exponential scaling with relative temperature (Fig. 3(b)): the x-ray data for $x = 0.16$, 0.2 and 0.33 scale with $(T - T_{LTO})$ after normalization by the intensity of a nearby Bragg peak. The corresponding neutron scattering data are excluded here, since detector saturation for nearby Bragg peaks prevented accurate normalization. As demonstrated in Fig. 3(c), when the correlation length data of Fig. 2(c) are plotted versus $T - T_{LTO}$, they are consistent with power-law scaling at low relative temperatures, albeit with a somewhat smaller exponent than the prior result of ~1/3 for LSCO [37,39]. We note that linear extrapolation of $T_{LTO}(x)$ to effective negative temperatures was used for $x = 0.24$ and 0.33 [37]. The scaling breaks down at high relative temperatures, where the characteristic length approaches 2-3 in-plane lattice constants. This breakdown is not surprising, since non-universal atomic details are expected to become relevant as the characteristic length approaches a single in-plane lattice constant. Moreover, the data for $x = 0.24$ seem to follow a somewhat different trend than the other samples (including both higher and lower Sr concentrations) and are closer aligned with the results for LSCO. The reason for this difference is unclear, but we note that, for $x = 0.24$, LSNO shows a particularly large discrepancy between the exponential decay rates of energy-integrated and quasi-elastic LTO intensities (Fig. 2(e) and Fig. 8 in Appendix C), which could be related to this dichotomy.

We also investigate spin and charge fluctuations above their respective bulk ordering temperatures, $T_{SO}$ and $T_{CO}$ (Fig. 4). Studies of the spin and charge response of LSNO have revealed an incommensurate 'stripe' phase characterized by nanoscale separation between hole-rich and hole-poor regions [8-14]. Our data are consistent with these studies: incommensurate spin- and charge-density-wave fluctuations appear together and become stronger as the Sr concentration approaches 1/3. Moreover, the observed spin fluctuations are essentially uncorrelated perpendicular to the Cu-O planes (see Fig. 7 in Appendix B), signifying their two-dimensional nature, whereas the charge fluctuations show an observable modulation along $L$, but the out-of-plane correlation lengths are still limited to a few interplane distances. Our measurements using the CORELLI spectrometer are of sufficiently high fidelity to enable quantitative determination of the intensity of the incommensurate spin fluctuation peaks in a wide temperature range, while we use x-ray data to analyze the charge sector due to better signal-to-noise ratios. Both the magnetic and charge fluctuations follow exponential behavior (see also Figs. 9 and 10 in Appendix D). For $x = 0.16$, the magnetic and structural total scattering ($T_0 = 147 \pm 7$ K vs. $256 \pm 20$ K) and quasistatic scattering ($182 \pm 33$ K vs. $147 \pm 11$ K) decay constants fall into the $200 \pm 60$ K range. For $x = 0.24$, the magnetic total scattering decay constant is comparable to the structural value ($345 \pm 18$ K vs. $303 \pm 18$ K), whereas the quasistatic magnetic scattering decreases much more gradually ($714 \pm 153$ K vs. $130 \pm 8$ K). Regarding the charge response (measured in detail for $x = 0.20$), we find that $T_0$ is about a factor of two larger than the corresponding structural value ($435 \pm 17$ K vs. $217 \pm 4$ K). Notably, since the CO intensities are obtained from x-ray measurements, it is not possible to determine the relative weights of static and dynamic components (unlike in the case of the structural and magnetic neutron data in Figs. 2(d,e) and 4(b)). In principle, it is possible that the observed CO temperature dependence originates from a redistribution of spectral weight, but we believe it to be unlikely for such processes to yield an exponential temperature dependence over an extended range and with a decay rate that is comparable to those obtained for the structural and magnetic degrees of freedom.

## IV. Discussion

The extended exponential decay, now also uncovered in the charge and magnetic degrees of freedom of LSNO, and the scaling with relative (rather than reduced) temperature are both inconsistent with critical fluctuations associated with a second-order thermal phase transition. In LSCO, the unconventional scaling, observed in both LTO and superconducting fluctuations, was interpreted as evidence for structural rare-region effects within Landau-Ginzburg-Wilson theory beyond mean-field [40-43]. Although this theory is typically applied in the context of magnetism the general observations are, in principle, applicable to any system with rare-region physics. In the presence of exponentially rare ordered regions, it is predicted that the observable strength of the ordered phase should, to leading order, vary exponentially above the bulk transition temperature, $T_{cr}$, and that the relevant measure of the distance from the bulk transition is the relative temperature, $T - T_{cr}$, rather than the reduced temperature, $(T - T_{cr})/T_{cr}$ [40]. Our observations are thus indicative of a rare-region effects in the structural, charge and magnetic degrees of freedom of LSNO. Notably, the respective phase transitions might be weakly first-order, in which case the underlying mean-field transition temperature $T_{cr}$ would be somewhat lower than the experimentally determined values.

As noted, the characteristic lengths in LSNO and LSCO follow power-law behavior with somewhat different exponents. In both cases, the exponent is smaller than the mean-field value of 1/2 that is theoretically expected for the size of the rare ordered regions above the transition [40]. If the HTT-to-LTO transition is weakly first-order, as indicated by NMR and neutron scattering results for LSCO [53,54], an effective transition temperature $T'_{LTO} < T_{LTO}$ should be used. For LSCO, it was observed that the data approximately follow a power-law with exponent 1/2 if $T'_{LTO} = T_{LTO} - 25$ K [37]. For this to hold in the case of LSNO, an effective $T'_{LTO} = T_{LTO} - 120$ K is required (see Fig. 11 in Appendix D). Such a large shift seems unphysical, only results in approximate scaling over a limited temperature range, and this issue will require further investigation.

What might cause the underlying inhomogeneity? We note that: (i) the structural scaling behavior also holds for the undoped parent compound $La_2CuO_4$, which is free of doping-related point disorder, and to temperatures approaching 1000 K [37]; (ii) the superconducting scaling behavior also holds for compounds such as simple tetragonal $HgBa_2CuO_{4+\delta}$, which shows no LTO correlations [32-35]; (iii) both the perovskite $SrTiO_3$ and tetragonal $Sr_2RuO_4$, which is isostructural to HTT LSCO and LSNO, exhibit superconducting fluctuations that decay exponentially with temperature [35]. The totality of these results points to inherent "hidden" structural inhomogeneity and closely related emergent electronic behavior in perovskites and related materials. In this picture, the free energy of the system is lowered because of correlated nano- or meso-scale deviations from average structure, and the coupling of structural and electronic order parameters to the associated local strain field results in rare-region physics. We emphasize, however, that the observed scaling likely represents emergent behavior, which does not depend on the details of the underlying, microscopic inhomogeneity – the latter is only required to be present, but its nature might differ for different compounds and materials families. In perovskite-derived lamellar materials (such as LSNO), the atomic sizes in the perovskite and rock-salt layers are slightly mismatched for the ideal tetragonal structure, and Jahn-Teller distortions (e.g., of the $NiO_6$ octahedra in LSNO) can appear as well. The interplay of these effects lies at the origin of the LTO transition [55,56], but could also lead to more subtle short-range local strains that minimize the overall elastic energy. This is certainly the case for substitutionally-doped compounds such as LSNO, where the associated La/Sr point disorder can induce correlated displacements and lattice strains. However, in principle, local distortions can arise spontaneously and independently of chemical inhomogeneity, analogous to, *e.g.*, self-organized mesoscale structures associated with martensitic transitions [24]. Since the relevant length scales in LSNO and LSCO are relatively small, we expect large-scale numerical simulations of the structural degrees of freedom to be feasible and capable of providing further insight into the free energy landscape.

## V. Summary

To summarize, we find unusual scaling and exponentially decaying nanoscale fluctuations of octahedral tilts as well as spin- and charge-density waves across the temperature-doping phase diagram of the lamellar nickelate $La_{2-x}Sr_xNiO_4$. This behavior is identical to that recently reported for orthorhombic fluctuations in several cuprates and for superconducting fluctuations in cuprates and other complex oxides. These findings imply that inherent, correlated structural inhomogeneity is also present in the nickelates and pivotal to their emergent electronic behavior, including superconductivity.

## Acknowledgements

This work was supported by the U.S. Department of Energy (DOE) through the University of Minnesota Center for Quantum Materials, under Grant No. DE-SC-0016371, by the Croatian Science Foundation under Grant No. UIP-2020-02-9494, and the Croatian Ministry of Science, Education and Youth. Work at Argonne (R.O.) was supported by the DOE Office of Science, Basic Energy Sciences, Materials Sciences and Engineering Division. A portion of this research was performed on APS beam time award GUP - 807978 from the Advanced Photon Source, a U.S. DOE Office of Science user facility operated for the DOE Office of Science by Argonne National Laboratory under Contract No. DE-AC02-06CH11357. A portion of this research used resources at the Spallation Neutron Source, a DOE Office of Science User Facility operated by the Oak Ridge National Laboratory, on proposal numbers IPTS-30991 and IPTS-33325.

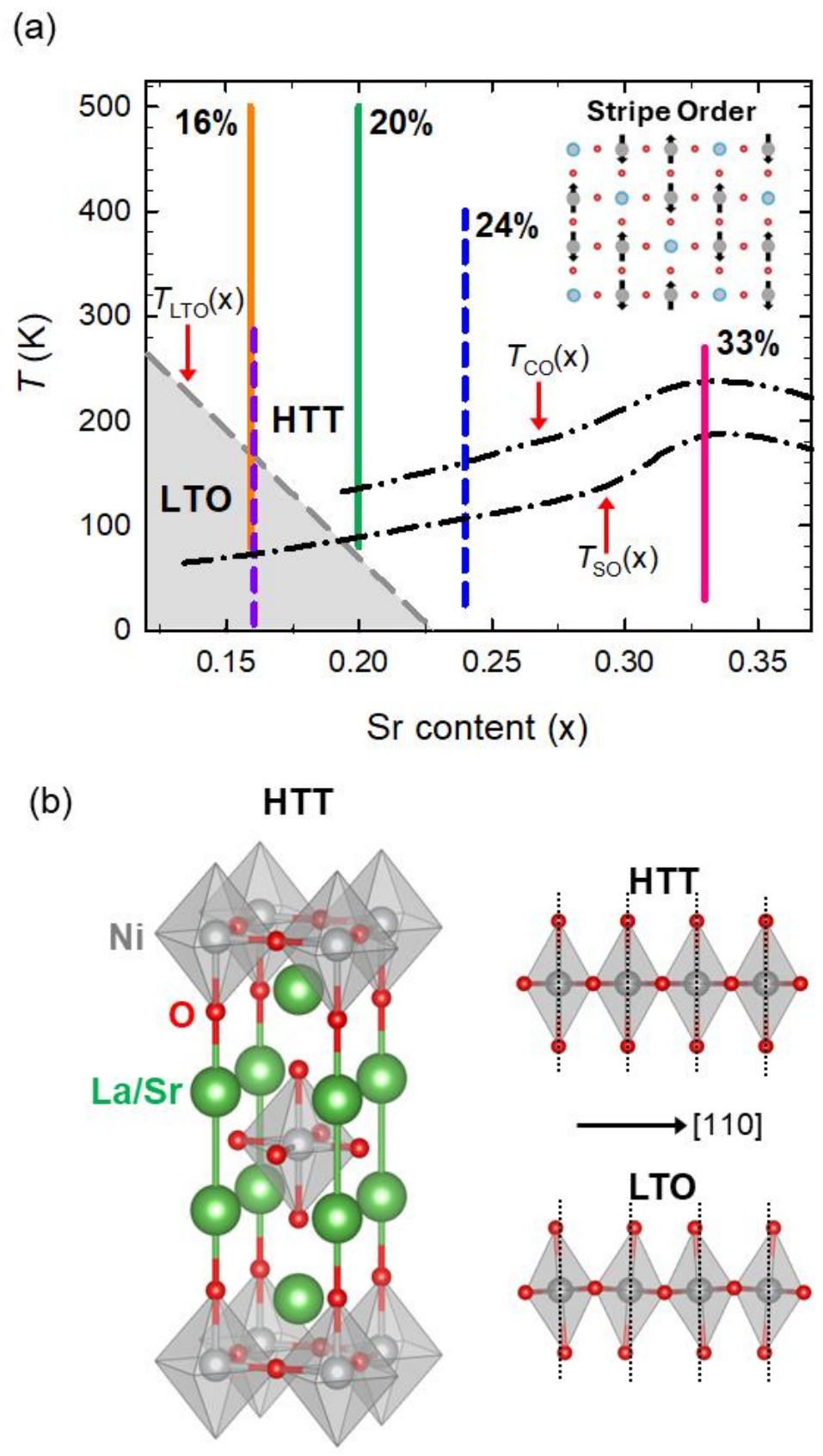

FIG. 1: Structural, charge, and magnetic order in $La_{2-x}Sr_xNiO_4$ (LSNO). Temperature-doping phase diagram (HTT/LTO phase boundary from [7] and present work; SO and CO phase boundaries from [10,48]). $T_{LTO}(x)$, $T_{CO}(x)$ and $T_{SO}(x)$ are the structural, charge, and spin transition temperatures, respectively. Solid vertical lines: X-ray data ($x$ = 0.16, 0.20 and 0.33). Dashed vertical lines: neutron data ($x$ = 0.16 and 0.24). $T_{LTO}(x)$ is nearly the same as for $La_{2-x}Sr_xCuO_4$ (LSCO). Inset: schematic of correlated charge-spin stripe order in the $NiO_2$ planes for $x$ = 0.33. Gray and red circles: Ni and O atoms, respectively. Blue circles: doped holes. Neighboring spins (arrows) order antiferromagnetically; hole-rich charge stripes form anti-phase domain walls. **(b)** High-temperature tetragonal (HTT) unit cell and the associated distortion leading to the low-temperature orthorhombic (LTO) structural phase. In the LTO phase, the $NiO_6$ octahedra ($CuO_6$ octahedra in LSCO) rotate about either [110] or [$1\bar{1}0$], leading to two types of domains.

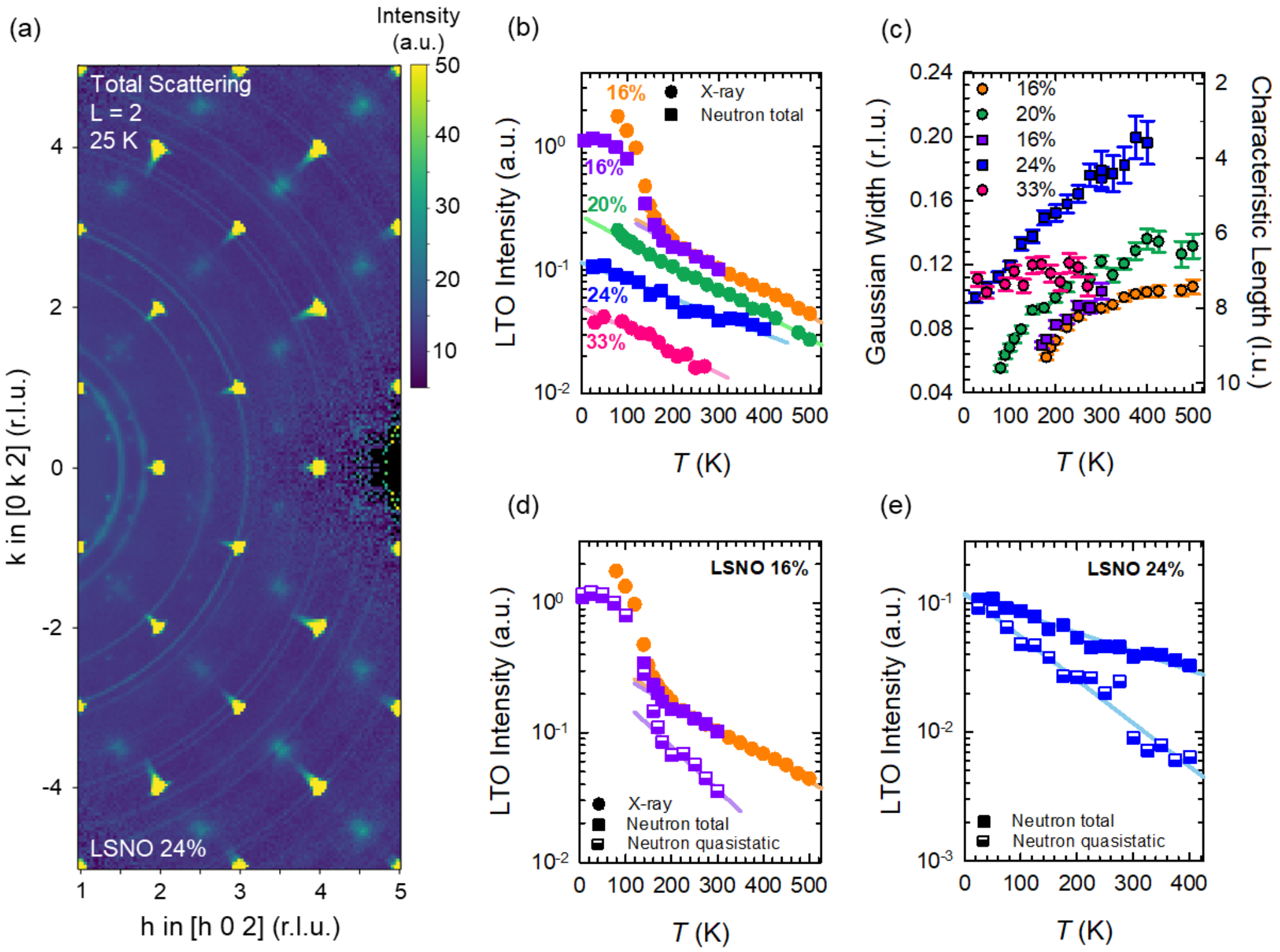

FIG. 2: Orthorhombic fluctuations in the tetragonal phase. (a) Neutron total scattering intensity in the *HK*2 plane for $x = 0.24$ at 25 K. Diffuse LTO fluctuations are clearly seen at half-integer positions. (b) Semi-log plot of LTO intensity vs. temperature reveals an exponential decay in the HTT phase in both x-ray and neutron total scattering data; the characteristic decay rate ($T_0$) is essentially doping-independent. (c) Gaussian widths/characteristic lengths, extracted from fits to the LTO peaks. (d,e) Comparison of total and quasistatic neutron scattering LTO intensity for $x = 0.16$ (d) and $x = 0.24$ (e) reveals distinct exponential decay rates.

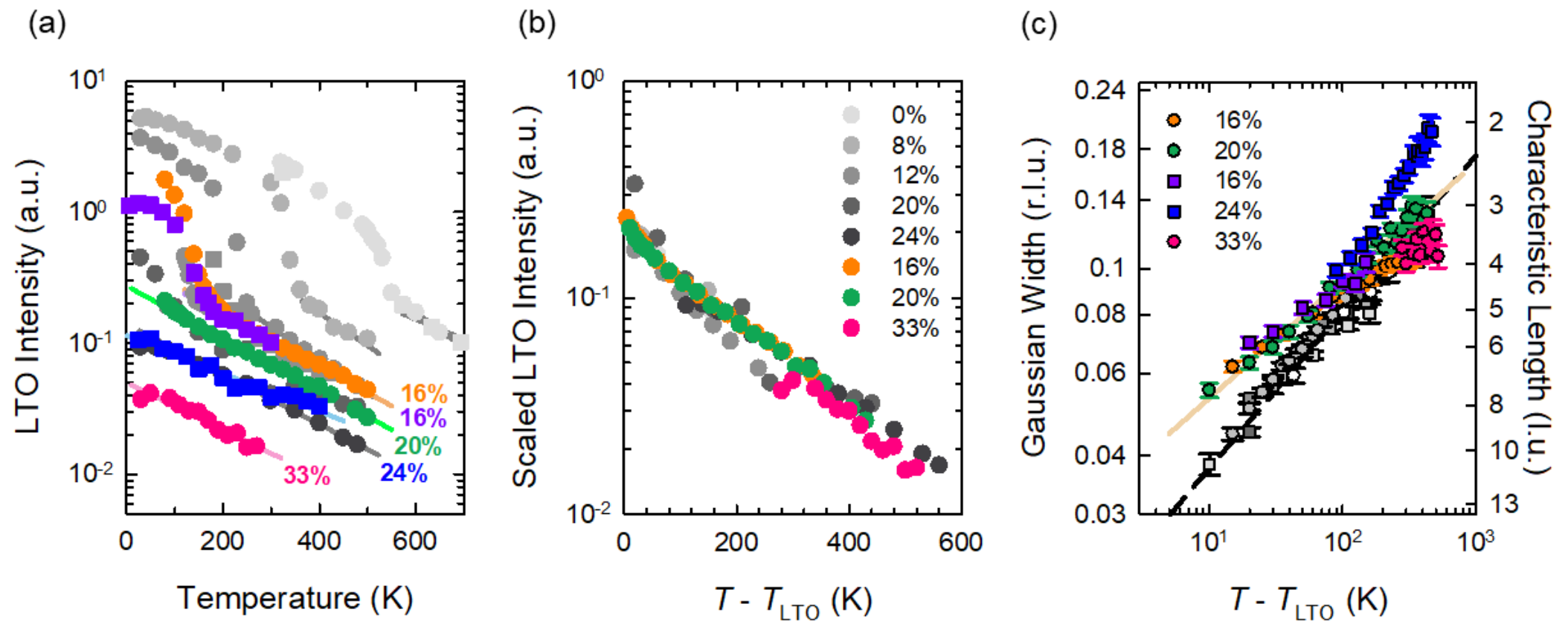

FIG. 3: (a) Comparison of LTO intensity (same data and colored symbols as in Fig. 2(b)) with prior x-ray results [37] for $La_{2-x}Sr_xCuO_4$ (LSCO, same gray circles for $x$ = 0 to 0.24 as in (b)) reveals closely similar exponential decay rates ($T_0$ ~ 200 K to 300 K) above $T_{LTO}(x)$. (b) Exponential scaling of x-ray LTO intensity with relative temperature, $T$-$T_{LTO}$, with intensities normalized to those of a nearby Bragg peak at 80 K (90 K for $x$ = 0.33). All three LSNO data sets were multiplied by the same factor for comparison with the LSCO data. (c) Power-law scaling of gaussian widths/characteristic lengths with relative temperature (same data and symbols as in Fig. 2(c)) and comparison with prior results for LSCO [25]. The LSNO data approximately scale for lengths larger than ~4 l.u., consistent with an exponent of ~1/4 (solid line), whereas the LSCO data follow a power-law with exponent ~1/3 (dashed line).

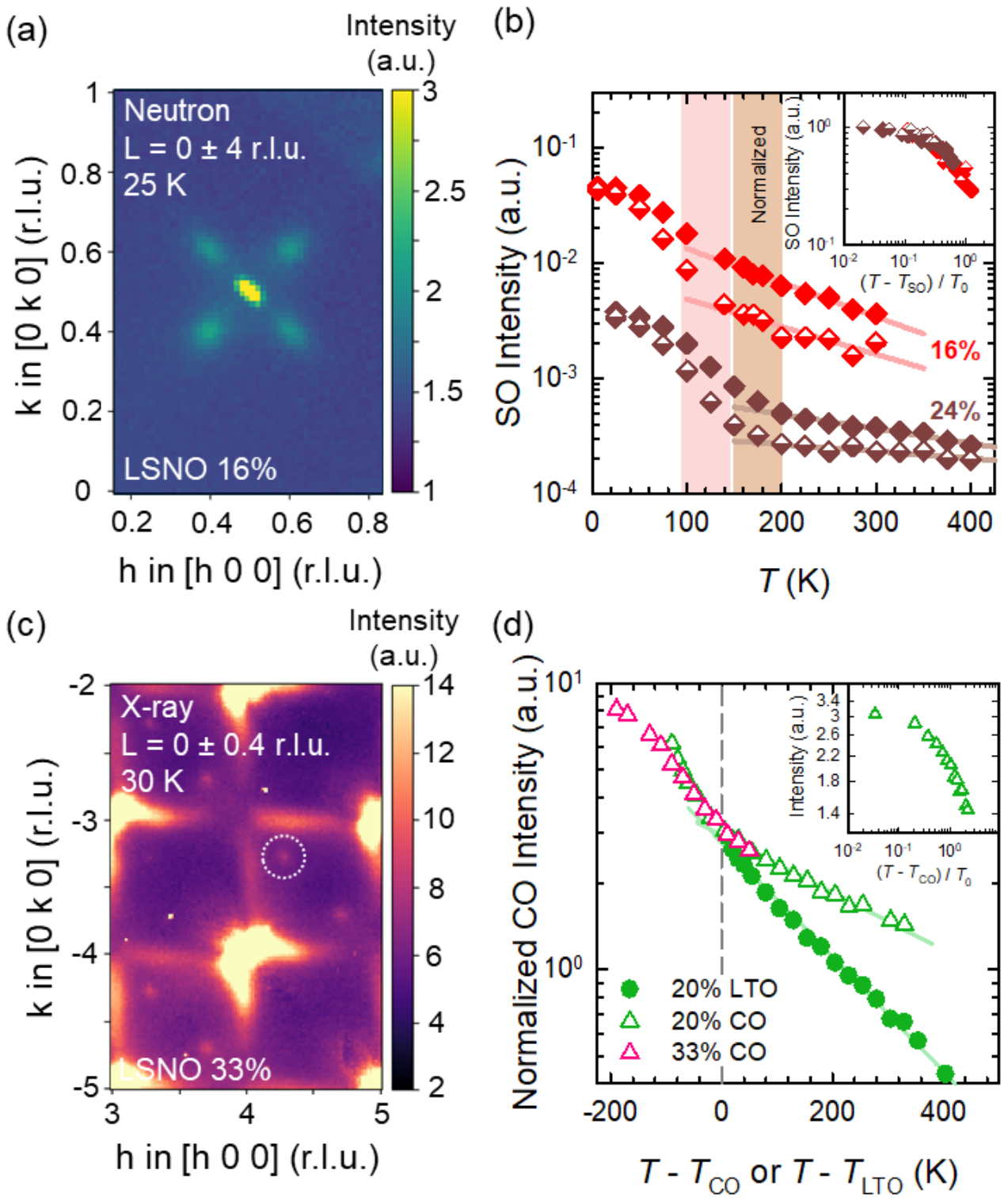

Fig. 4: Magnetic and charge fluctuations in LSNO observed using neutrons and x-rays, respectively. (a) Total neutron scattering data in the *HK*-plane (*L*-integration from -4 to 4 r.l.u.) for $x = 0.16$ at 25 K. Incommensurate magnetic peaks are seen at ($0.5\pm\delta$ $0.5\pm\delta$ $L$). (b) Spin-order (SO) response for $x = 0.16$ and $x = 0.24$ (on a semi-log scale) exhibits exponential behavior in both the quasistatic and total scattering channels. Inset: log-log plot of normalized SO intensity vs. relative temperature ($T - T_{SO}$, normalized by the respective characteristic temperature scale, $T_0$): the response is inconsistent with power-law behavior (see also Fig. 9 in Appendix D). The shaded pink and brown regions indicate the SO transition temperature range for the 16% and 24% samples, respectively. (c) X-ray data for LNSO 33% in the *HK*0 plane showing incommensurate charge order (CO) peaks (dashed circle: representative CO peak). (d) CO intensity for $x =$ 0.20 and 0.33 vs. relative temperature, $T - T_{CO}$, on a semi-log scale. The intensities were normalized to overlap for visual clarity. Detector artifacts in the $L = 0$ plane prevented proper normalization by nearby Bragg peak intensities. The LTO response for $x = 0.20$ is shown for comparison. The vertical dashed line indicates the respective transition temperature. The CO fluctuations exhibit exponential behavior. Inset: log-log plot of CO intensity vs. relative temperature, normalized by $T_0$: the data are inconsistent with power-law behavior (see Fig. 10 in Appendix D).

**Appendix A: Intensity and Characteristic Length Extraction**

This Appendix shows typical raw data and fits used to extract the intensity and characteristic length temperature dependences. Figure 5 shows the raw intensity, background, and background-subtracted results for representative LTO, CO, and SO features. Figure 6 shows representative line cuts through LTO features and corresponding gaussian fits after subtracting the fit third-order polynomial background.

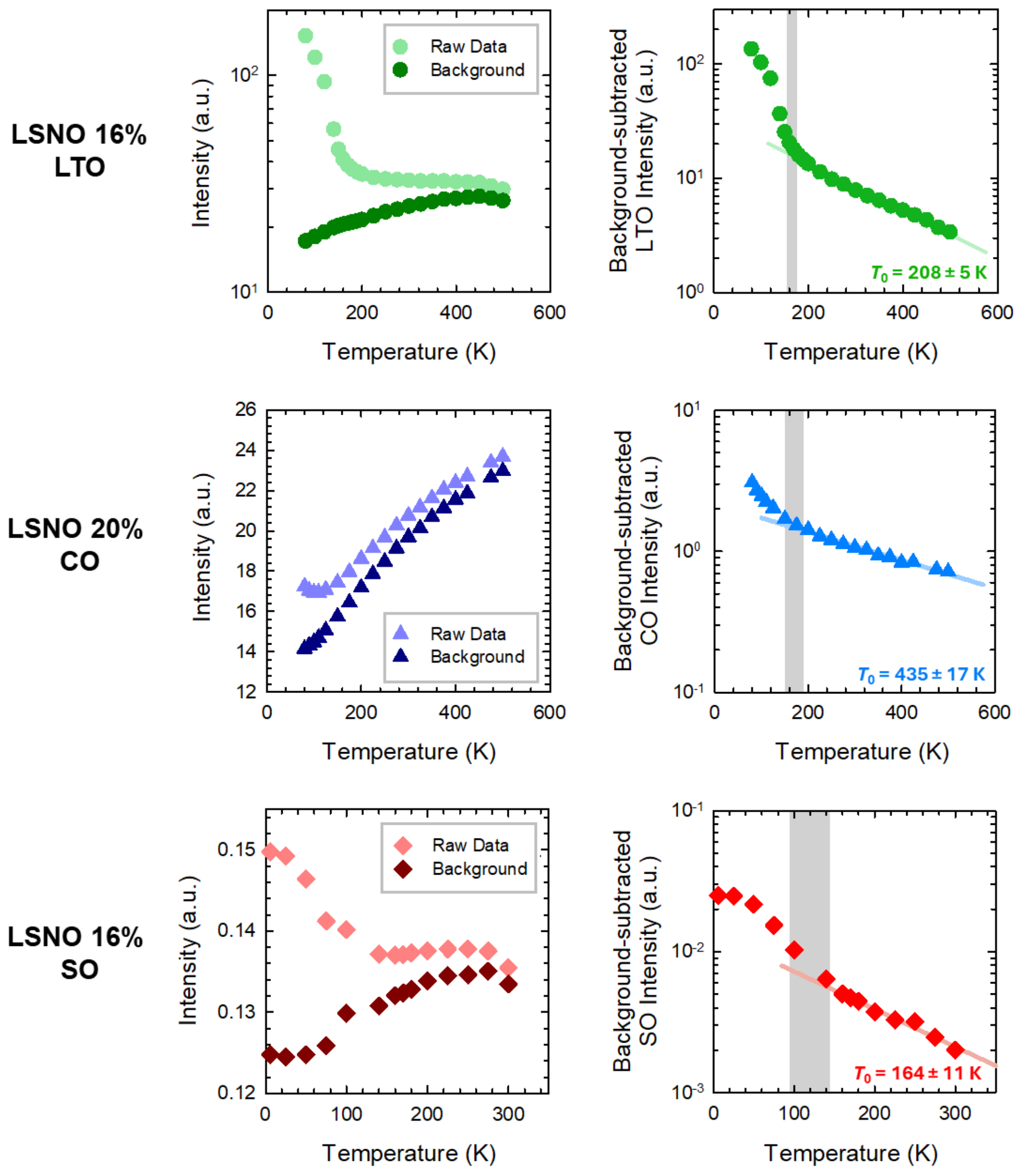

Fig. 5: Temperature dependence of raw data, background, and background-subtracted intensity for LTO fluctuations in LSNO 16% (x-rays, first row), CO fluctuations in LSNO 20% (x-rays, second row), and SO fluctuations in LSNO 16% (neutron total scattering, third row). Exponential fits and characteristic temperatures are indicated for the background-subtracted intensities. Shaded gray regions indicate transition temperature estimates.

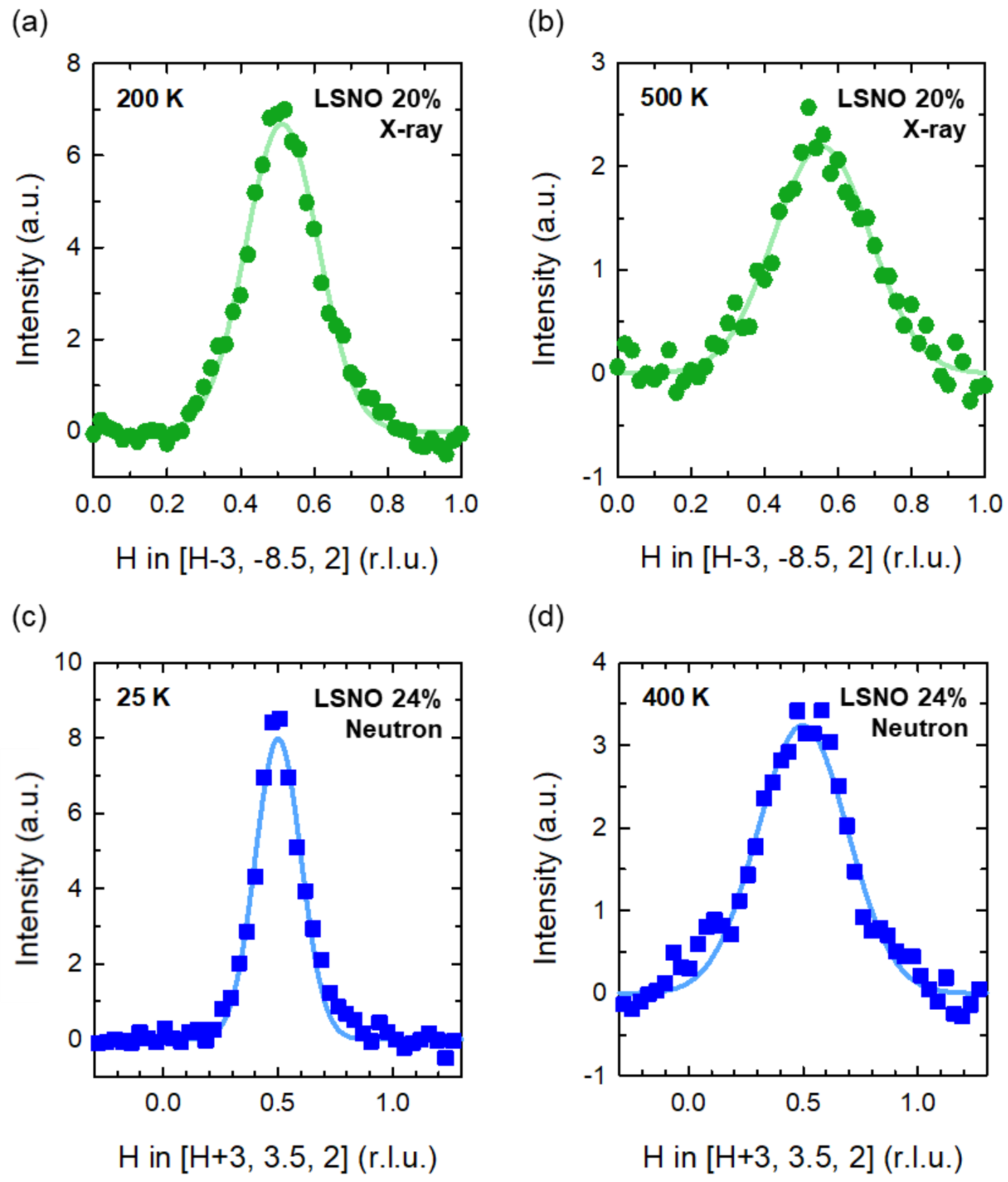

Fig. 6: Background-subtracted momentum cuts along [100], with intensity integrated along [010] and [001]. (a,b) X-ray data for LSNO 20% at (a) 200 K and (b) 500 K. (c,d) Neutron total scattering data for LSNO 24% at (c) 25 K and (d) 400 K. The background was modelled as a third-order polynomial. Solid lines: Gaussian fits. The extracted Gaussian widths were used to calculate the characteristic length scales shown in Fig. 3(c).

## Appendix B: Spin Fluctuations

The spin fluctuations form rods of diffuse scattering along [001], indicative of planar correlations. This Appendix shows representative data.

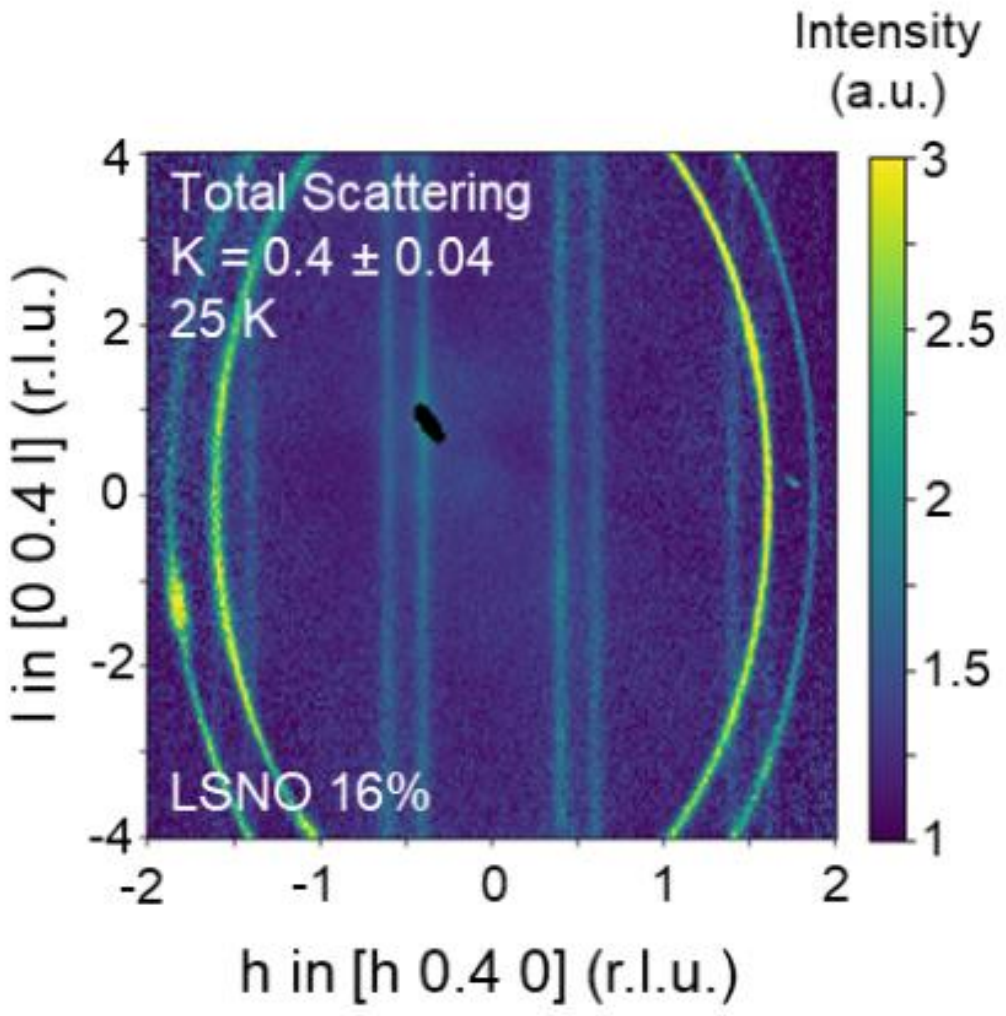

Fig. 7: Total neutron scattering in the *H*0.4*L* plane (after *K*-integration from 0.36 to 0.44 r.l.u.) for LSNO $x$ = 0.16 at 25 K. The magnetic fluctuations form rods along [001], demonstrating the two-dimensional nature of the response.

## Appendix C: Temperature Scales

Summary of characteristic LTO, CO, and SO temperature scales for LSNO from this work and LSCO [37,39]. Figure 8 shows the extracted temperature scales as a function of strontium content. The characteristic temperatures are tabulated in Table 1.

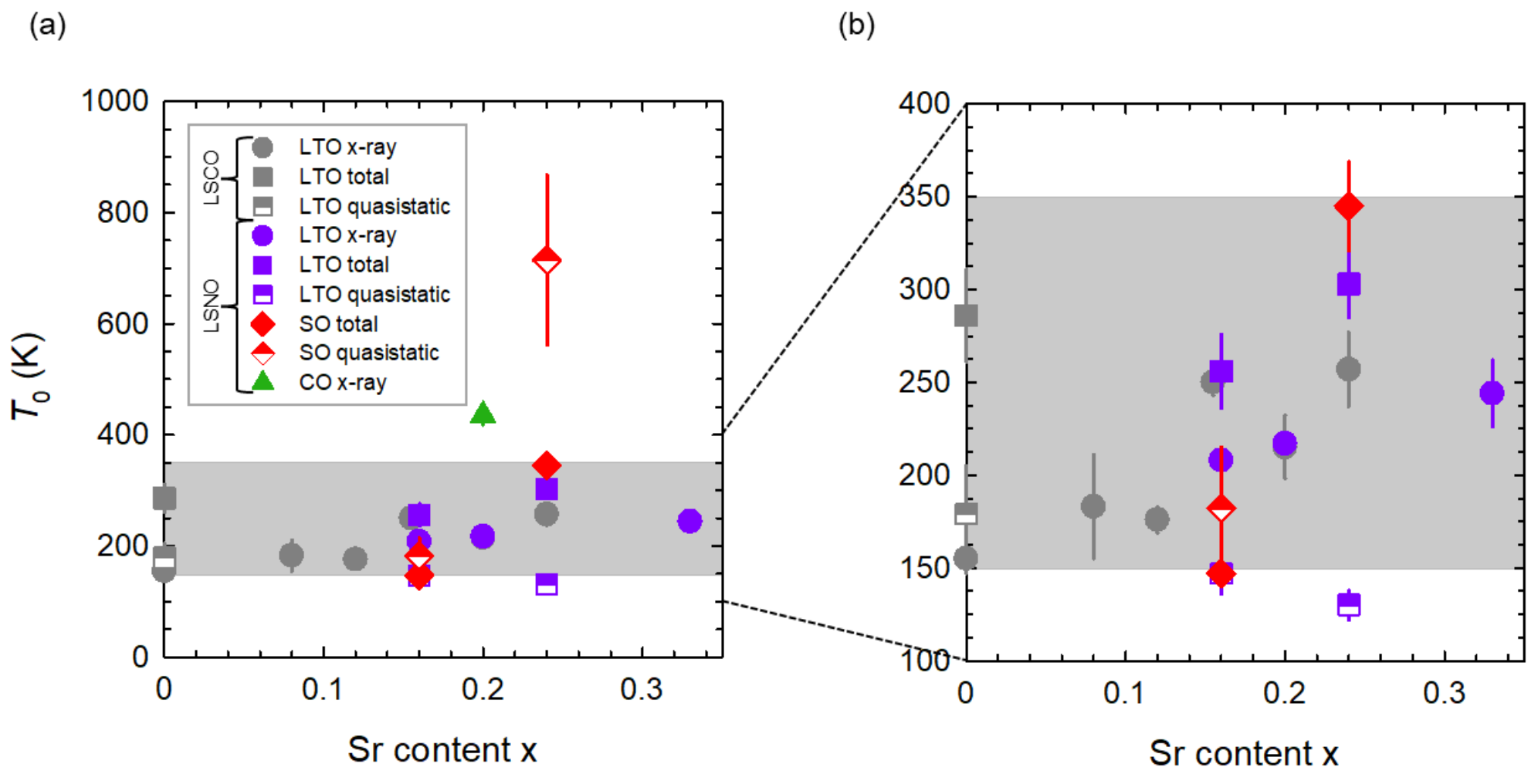

Fig. 8: (a) Characteristic temperature scales vs. strontium content. (b) Same as (a), but for a narrower temperature range. Nearly all $T_0$ values fall into the 150 K to 350 K range (gray band). This analysis does not account for possible subtle systematic errors introduced by the background subtraction.

| **Transition Metal** | **Sr content x** | **Feature** | **$T_c$ (K) (LTO, CO, SO)** | **$T_0$ (K) X-ray** | **$T_0$ (K) Neutron Total** | **$T_0$ (K) Neutron Quasistatic** |
|---|---|---|---|---|---|---|
| Cu | 0 | LTO | 535 ± 10 | 155 ± 8 | 286 ± 25 | 179 ± 26 |
| Cu | 0.08 | LTO | 345 ± 10 | 183 ± 28 | — | — |
| Cu | 0.12 | LTO | 235 ± 10 | 176 ± 7 | — | — |
| Cu | 0.155 | LTO | 145 ± 5 | 250 ± 7 | — | — |
| Ni | 0.16 | LTO | 165 ± 10 | 208 ± 5 | 256 ± 20 | 147 ± 11 |
| Ni | 0.16 | SO | 120 ± 25 | — | 147 ± 7 | 182 ± 33 |
| Cu | 0.2 | LTO | 25 ± 15 | 215 ± 17 | — | — |
| Ni | 0.2 | LTO | 70 ± 10 | 217 ± 4 K | — | — |
| Ni | 0.2 | CO | 170 ± 20 | 435 ± 17 | — | — |
| Cu | 0.24 | LTO | — | 257 ± 20 | — | — |
| Ni | 0.24 | LTO | — | — | 303 ± 18 | 130 ± 8 |
| Ni | 0.24 | SO | 175 ± 25 | — | 345 ± 24 | 714 ± 153 |
| Ni | 0.33 | LTO | — | 244 ± 18 | — | — |
| Ni | 0.33 | CO | 220 ± 10 | — | — | — |

Table 1: Comparison of LSNO characteristic orthorhombic (LTO), spin-order (SO), and charge-order (CO) decay rates with prior results for LSCO (shaded gray) [37,39].

### Appendix D: Temperature Dependences

Additional information pertaining to the observed SO intensity, CO intensity, and LTO characteristic length temperature dependences. For LSCO, the latter was previously considered in the Supplemental Material of [37]. Figures 9 and 10 show, respectively, the SO and CO intensity temperature dependences on semi-log, log-log, and linear scales. In both cases, the temperature dependence is most consistent with an exponential decay. Figure 11 compares the LTO characteristic length scaling with a power-law with an exponent of 1/2, which is inconsistent with the data.

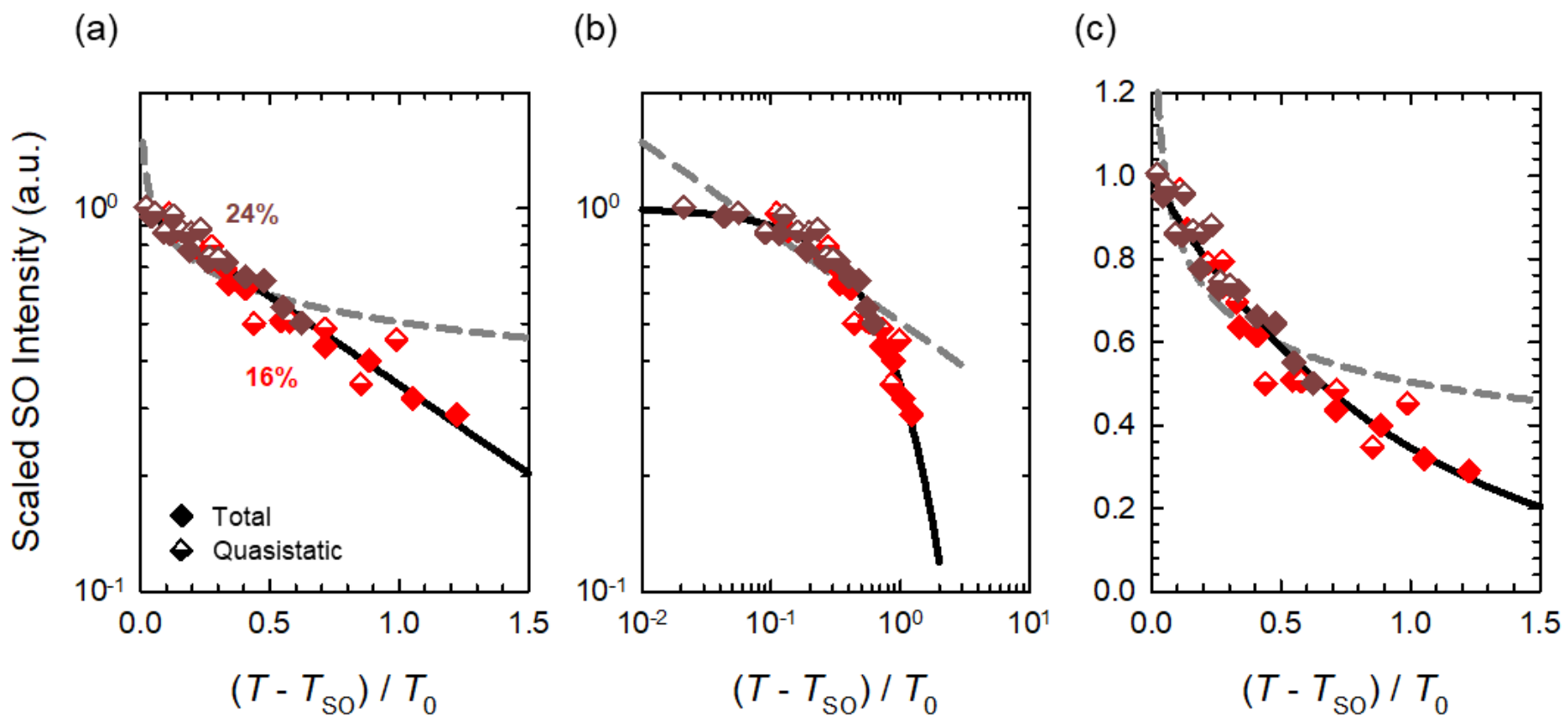

Fig. 9: Temperature dependence of the momentum-integrated SO neutron scattering intensity for LSNO $x = 0.16$ (red diamonds) and 0.24 (brown diamonds) on (a) semi-log, (b) log-log, and (c) linear scales. The relative temperatures are normalized by the respective temperature scales $T_0$ obtained from fits to ~$\exp(-T/T_0)$ (see Table 1 in Appendix C). The intensities are scaled for comparison. The data are consistent with exponential (solid lines) rather than power-law (dashed lines) behavior.

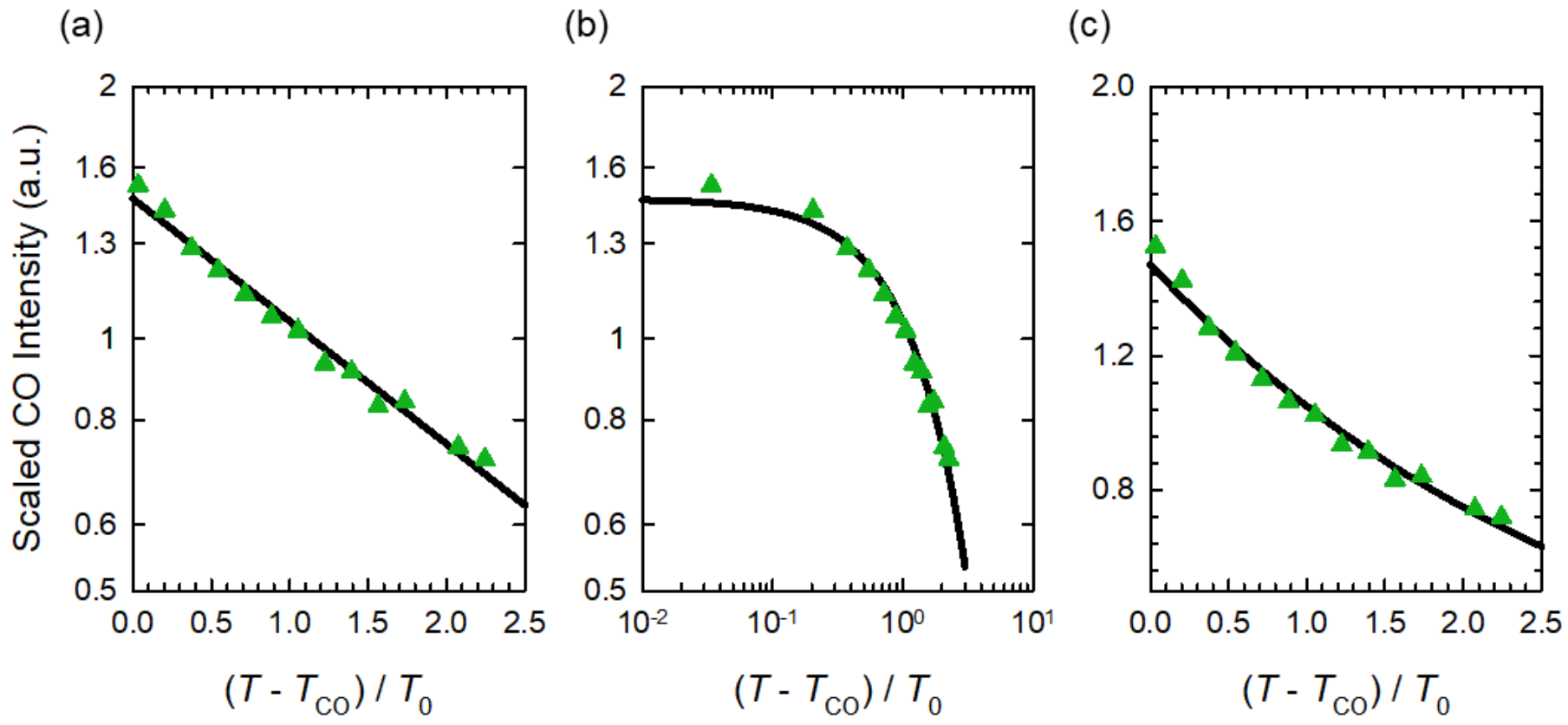

Fig. 10: Temperature dependence of the momentum-integrated x-ray scattering CO intensity for LSNO $x = 0.20$ on (a) semi-log, (b) log-log, and (c) linear scales. The relative temperature is normalized by the CO temperature scale $T_0 = 435$ K for this sample. The temperature dependence is consistent with exponential decay (solid lines) rather than power-law behavior.

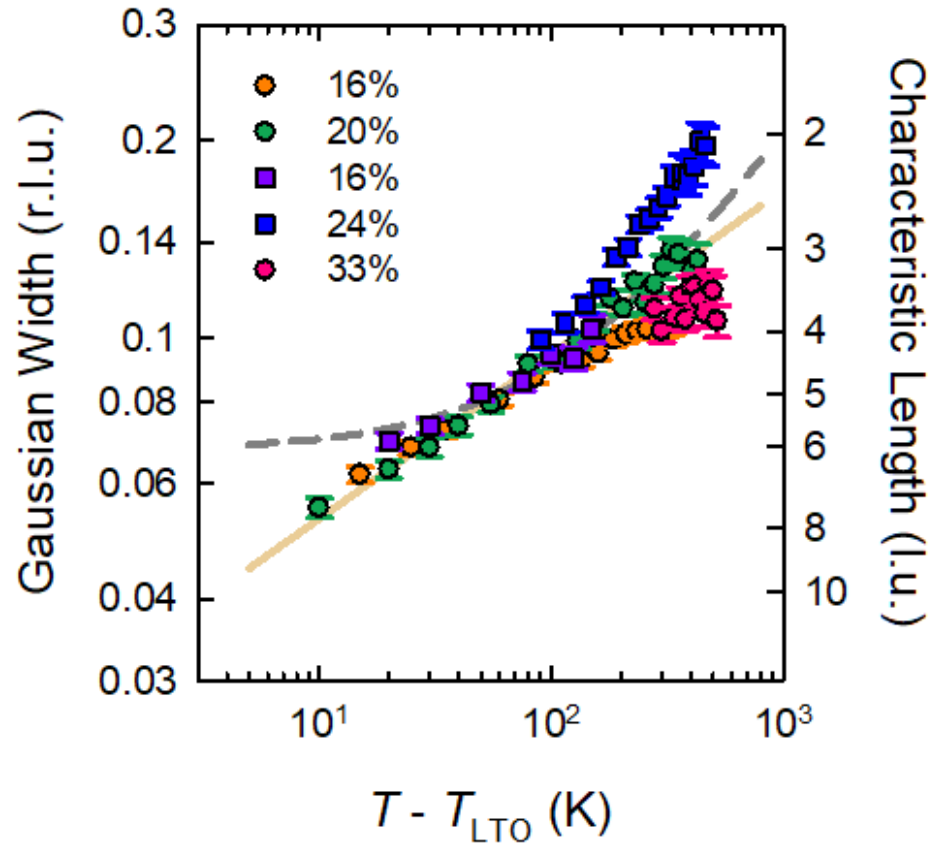

Fig. 11: Scaling of gaussian width and characteristic length with relative temperature. The solid line indicates a power-law with an exponent of 1/4, whereas the dashed line indicates a power-law with the expected mean-field exponent of 1/2 and with an effective $T'_{LTO} = T_{LTO} - 120$ K.